\documentclass[prl,a4paper,aps,showpacs,twocolumn,floats]{revtex4}
\usepackage{epsfig}

\newcommand{\boe}{{\mathbf e}}

\newcommand{\bn}{{\mathbf n}}
\newcommand{\bv}{{\mathbf v}}

\newcommand{\MM}{{\cal M}}

\newcommand{\cd}{\cdot}

\newcommand{\de}{\delta}
\newcommand{\De}{\Delta}

\newcommand{\La}{\Lambda}

\newcommand{\Om}{\Omega}

\newcommand{\be}{\begin{equation}}
\newcommand{\ee}{\end{equation}}
\newcommand{\gsim}{\stackrel{>}{\sim}}
\newcommand{\lsim}{\stackrel{<}{\sim}}
\newcommand{\bea}{\begin{eqnarray}}
\newcommand{\eea}{\end{eqnarray}}
\newcommand{\bean}{\begin{eqnarray*}}
\newcommand{\eean}{\end{eqnarray*}}

\newcommand{\mr}{\mathrm}

\newcommand{\LCDM}{\Lambda\rm{CDM}}

\newcommand{\HH}{{\ensuremath{\mathcal H}}}

\newcommand{\pz}{^{(0)}}
\newcommand{\pu}{^{(1)}}

\begin{document}

\title{Dipole of the luminosity distance: a direct measure of $H(z)$}

\author{Camille Bonvin}
\email{camille.bonvin@physics.unige.ch}
\affiliation{D\'epartement
de Physique Th\'eorique, Universit\'e de
  Gen\`eve, 24 quai Ernest Ansermet, CH--1211 Gen\`eve 4, Switzerland}

\author{Ruth Durrer}
\email{ruth.durrer@physics.unige.ch}
\affiliation{D\'epartement de
Physique Th\'eorique, Universit\'e de
Gen\`eve, 24 quai Ernest Ansermet, CH--1211 Gen\`eve 4, Switzerland}

\author{Martin Kunz}
\email{martin.kunz@physics.unige.ch}
\affiliation{D\'epartement de Physique Th\'eorique, Universit\'e
de Gen\`eve, 24 quai Ernest Ansermet, CH--1211 Gen\`eve 4,
Switzerland}

\date{\today}

\begin{abstract}
We show that the dipole of the luminosity distance is a useful
observational tool which allows us to determine the Hubble parameter as a
function of redshift $H(z)$. We determine the number of
supernovae needed to achieve a given precision for $H(z)$ and to
distinguish between different models for dark energy. We analyse a
sample of nearby supernovae and find a dipole consistent with the cosmic
microwave background at a significance of more than 2$\sigma$.
\end{abstract}


\maketitle


One of the biggest cosmological surprises in recent years was the
discovery that the Universe is presently undergoing a phase of accelerated
expansion~\cite{snIa}. The reason for this behaviour
is still a complete mystery.

If the Universe is homogeneous and
isotropic on large scales, all contributions to the
cosmological energy momentum tensor are characterised by their
energy density $\rho(z)$ and pressure $P(z)$ as functions of
cosmic redshift $z$. Accelerated expansion requires that overall
$\rho+3P <0$ today. This can be achieved by introducing a
so-called 'dark energy' component with a very negative pressure in
addition to the usual pressureless matter. One of the main
challenges of observational cosmology is to characterise the
properties of this dark energy. The homogeneous and isotropic
aspects of dark energy are completely determined by the
equation of state parameter $w_\mr{de}(z)\equiv P_\mr{de}(z)
/\rho_\mr{de}(z)$ which links its pressure and energy density. The
primary goal of observational dark energy studies  is to measure
the function $w_\mr{de}(z)$.

Current experiments probing the dark energy equation of state
measure luminosity distances to supernovae or the angular diameter
distance to the last scattering surface via the cosmic microwave background (CMB) peak
positions. These distances are linked to $w_\mr{de}(z)$ through a
double integration, which renders them rather insensitive to rapid
variations of the equation of state. The required modelling can
lead to strong biases that are difficult to detect and quantify
\cite{Blanchard}. A direct measurement of the Hubble parameter
$H(z)$ would facilitate the derivation of $w_\mr{de}(z)$ immensely
and allow for a more direct comparison with model predictions. As
an explicit illustration, let us consider a flat universe. The
Friedmann equations then yield the following relation between
$w_\mr{de}(z)$  and $H(z)$
\be
w_\mr{de}(z)=\frac{3}{2}\left(\frac{2/3H'(z)+H^2(z)}{H^2(z)-H_0^2\Om_m(1+z)^3}
\right)~. \ee 
Apart from $H(z)$ and its derivative only the
parameter combination $\Omega_m H_0^2$ appears, which will be
measured by the Planck satellite to an accuracy of about 1\%.

It is possible to obtain $H(z)$ by computing the numerical derivative
of the distance data~\cite{daly}, but the current data
leads to a very noisy result. In the future, radial baryon oscillation
measurements should be able to measure $H(z)$ directly. Here we propose
an alternative, completely independent method based on luminosity
distance measurements.
\vspace{3mm}


In a previous paper~\cite{BDG} we have considered the luminosity distance
$d_L$ as function of the source redshift $z$ and direction $\bn$. We have shown
that not only the direction-averaged luminosity distance,
\be
d_L^{(0)}(z) = \frac{1}{4\pi} \int d\Omega_{\bn} d_L(z,\bn)
= (1+z)\int_{0}^{z}\frac{dz'}{H(z')}~,
\ee
but also its directional dependence, $d_L(z,\bn)$ can be of cosmological
interest. The directional dependence can be expanded
in terms of spherical harmonics, leading to observable multipoles, $C_\ell(z)$.
In this {\em letter} we concentrate on the dipole (corresponding
to $\ell=1$),
\be\label{def.dip}
 d_L^{(1)}(z) = \frac{3}{4\pi}\int d\Omega_{\bn} (\bn\cdot\boe) d_L(z,\bn)~.
\ee
Here $\boe$ is a unit vector denoting the direction of the dipole and
$d_L^{(1)}(z)$ is its amplitude.

As we have discussed in~\cite{BDG} (see also~\cite{Sasaki}), for
$z\gsim 0.02$ the dipole is dominated by the peculiar velocity of the
observer for all redshifts. The lensing contribution to
$C_1$ is of the order of $10^{-9}$ while our peculiar motion induces a
dipole of $C_1\simeq 10^{-3}-10^{-6}$ for $z\lsim 2$. At high
$\ell$'s, i.e. small scales 
the lensing contribution dominates for $z\gsim 1$,
 but this does not  interfere with the
dipole as it averages to zero under the integration~(\ref{def.dip}).
Neglecting multipoles
higher than the dipole we can write the full luminosity distance as
\be\label{dL.dip}
d_L(z,\bn) =  d_L^{(0)}(z) +  d_L^{(1)}(z)(\bn\cdot\boe) ~.
\ee

To derive a formula for $d_L^{(1)}$ (more details are
found in~\cite{BDG}), we use the luminosity
distance to a source emitting photons at conformal time $\eta$ in an
unperturbed Friedmann universe, $d_L^{(0)} = (1+z)(\eta_0-\eta)$.
The motion of the observer gives rise to a Doppler effect which is
the dominant contribution to the dipole,
\be\label{eqv0}
d_L(\eta,\bn) =  d_L^{(0)}(\eta)\left[1 - (\bn\cdot\bv_0)\right]~,
\ee
where $\bv_0$ is our peculiar velocity.
However, conformal time $\eta$ is not an observable quantity, but the
source redshift, $z=\bar{z}(\eta)+\delta z$ is. Here
$\bar{z}(\eta)=1/a(\eta)-1$ is the unperturbed redshift.
To first order
\be
 {d}_L(\eta,\bn) =
 {d}_L(\bar z,\bn) =  {d}_L(z,\bn) - \frac{d}{d\bar z}
 {d}_L^{(0)}(\bar z)\de z ~.
\ee
With $d_L^{(0)}(\bar z)=(1+\bar z)(\eta_0-\eta)$, we have
\bea
  \frac{d}{d\bar z} {d}_L^{(0)} &=& (1+\bar z)^{-1}
  d_L^{(0)} + \HH^{-1}(\bar{z})  \qquad \mbox{and } \nonumber \\
\de z &=& -(1+\bar z)(\bv_0\cd\bn) +  \mbox{ higher multipoles} ~.
\eea
Here $\HH(z) = H(z)/(1+z)$ is the co-moving Hubble parameter.
Inserting this in Eq.~(\ref{eqv0}), we obtain
\be\label{dip.H}
 d_L^{(1)}(z)(\bn\cdot\boe) = \frac{1+z}{\HH(z)}(\bn\cdot\bv_0)~.
\ee
Although $\bv_0$ is in principle a random variable, we can measure it directly
from the CMB dipole which is due to the same motion. Its magnitude is
$|\bv_0| = (368\pm2) \textrm{km/s}$ according to the COBE satellite
measurements \cite{CMBdip}. The amplitude of the luminosity distance dipole
is then
\be \label{dipamp}
d_L^{(1)}(z) = \frac{|\bv_0|(1+z)}{\HH(z)} = \frac{|\bv_0| (1+z)^2}{H(z)} ~,
\ee
and its direction is $\boe =\bv_0/|\bv_0|$.
The dipole in the  supernova data gives therefore a direct
measure of $H(z)$.
\vspace{2mm}

\begin{figure}[ht]
\centerline{\epsfig{figure=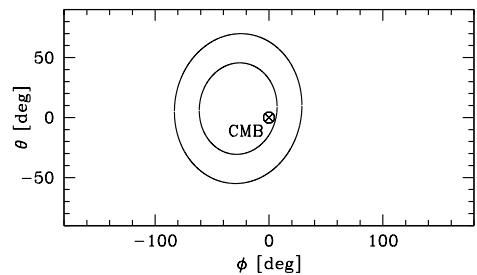,width=6.5cm}}
\caption{ \label{fig:dipole} Direction of the luminosity dipole from a
low-redshift supernovae sample, in a celestial coordinate system centred on the
CMB dipole ($1$ and $2\sigma$ contours). The two directions
agree well.
}
\end{figure}

As a first step we want to test whether there is a dipole present at
all in the supernova distribution, and if its direction and magnitude
is compatible with expression (\ref{dip.H}) (see also \cite{riess}). 
Supernova data is conventionally
quoted in terms of magnitudes rather than the luminosity distance.
The link between magnitudes and the luminosity distance is given by
\be\label{def.mag}
m-\MM = 5\log_{10}(d_L/10 \textrm{pc})
\ee
where $\MM$ is the intrinsic magnitude (however,
the absolute magnitude normalisation is degenerate with $\log(H_0)$
and is usually marginalised over). We use the low-redshift sample
of 44 supernovae assembled by the SNLS team \cite{snls}, together with
the supernova   locations from \cite{snlist}. To the given photometric
error we add an error for the peculiar velocity of the source of 300 km/s and a constant dispersion of
$\Delta m = 0.12$. The latter error ensures a reasonable goodness-of-fit
of both the monopole and dipole term. We subtract the monopole
of $m(z,\bn)$ and find the best fit value of $\bv_0$ for the
dipole. In Fig.~\ref{fig:dipole} we show the angular uncertainty of $\bv_0$.
The direction is compatible with the CMB dipole at
the $1\sigma$ level. The magnitude of the luminosity dipole gives
$|\mathbf{v}_0|=405\pm192$ km/s,
in good agreement with the CMB dipole value of $368$ km/s.

Fixing the CMB dipole direction and fitting only the amplitude, we
obtain $|\bv_0| =358$ km/s with $\chi^2_{\min} = 48.2$ whereas $\bv_0=0$
gives $\chi^2 = 52.7$. The absence of a dipole is therefore disfavored
at over $2\sigma$.  

Let us estimate the accuracy with which
we can determine $H(z)$ from a measurement of $N$
supernovae in a redshift bin $[z-dz,z+dz]$. We assume that the
magnitude of each supernova is known with a precision $\Delta m$
(independent of $z$). We consider $\bv_0$ given by the CMB
dipole and work with the ansatz (\ref{dL.dip}). The error in the
magnitude translates into an error on the luminosity distance,
\be\label{demag}
\de d_L(z,\bn)=\frac{\ln(10)}{5}d_L(z,\bn)\de m(z,\bn)~.
\ee
We add the error into our ansatz, setting
\bea
m(z,\bn) &=& m^{(0)}(z) +m^{(1)}(z)(\bn\cdot\boe) + \de m(z,\bn) \\
d_L(z,\bn) &=& d_L^{(0)}(z) +d_L^{(1)}(z)(\bn\cdot\boe) + \de d_L(z,\bn) .
\eea
We assume that different supernovae are uncorrelated, so that the
variance of the magnitude is given by
\be
\langle \de m(z,\bn) \de m(z,\bn')\rangle = 4\pi(\De m)^2\de^2(\bn-\bn') .
\ee
The error on the dipole can now be computed using
\be
\de d_L\pu(z)=\frac{3}{4\pi}\int d\Om_\bn (\bn\cdot\boe) \de d_L(z,\bn)~.
\ee
Its variance is
\bea
&& \left(\Delta d_L\pu(z)\right)^2=\Big\langle \left(\de
d_L\pu(z)\right)^2\Big\rangle\nonumber\\  \qquad &&
= 3\left(\frac{\ln(10)}{5}\right)^2 \Delta m^2
 \left(d_L\pz(z)\right)^2
~. \nonumber\\
\eea
 As the monopole is much larger than the dipole we have neglected the
 latter in the previous expression 
and obtain our final formula for the variance of the dipole
\be
\Delta d_L\pu(z)\simeq\frac{\sqrt{3}\ln(10)}{5} d_L\pz(z)\Delta m = 
 \sqrt{3}\Delta d_L(z)~.
\ee

The absolute error on the dipole is therefore comparable to the error on the
monopole and, not surprisingly, the relative error is much larger,
\be
\frac{\De d_L\pu(z)}{d_L\pu(z)} = \sqrt{3}\frac{\De d_L\pz(z)}{d_L\pu(z)}\gg
  \frac{\De d_L\pz(z)}{d_L\pz(z)}~.
\ee
 We will therefore need a large number of supernovae to determine
$H(z)$ with reasonable accuracy.

As the Hubble parameter is inversely proportional to the
dipole, its relative error is simply
\be
\frac{\Delta H(z)}{H(z)}=\frac{\Delta d_L\pu(z)}{d_L\pu(z)}=
\frac{\sqrt{3}\ln(10)}{5 |\mathbf{v}_0|} 
\frac{d_L\pz(z)H(z)}{(1+z)^2}\Delta m .
\label{eq:errorH}
\ee

\begin{figure}[ht]
\centerline{\epsfig{figure=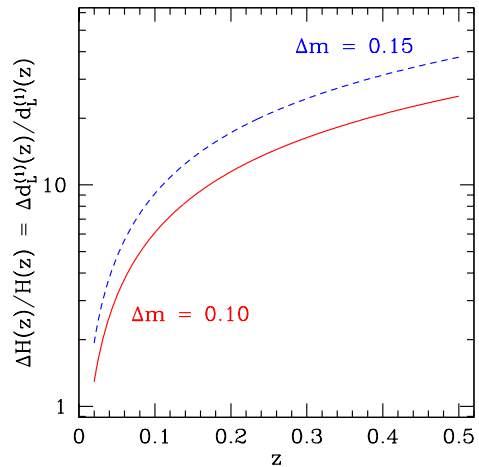,width=6.5cm}}
\caption{ \label{fig:H} We show the relative error in $H(z)$ for one
  supernova, as a function 
  of the redshift, in a flat $\Lambda$CDM universe with
  $\Omega_\La=0.7$ and for two different 
  values of the intrinsic error $\De m=0.1$ and $\De m=0.15$. This
  represents as well the relative error in the dipole $d_L\pu(z)$.
  }
\end{figure}

This formula is valid for any model of dark energy. Once we have measured the luminosity distance,
we can calculate the monopole and the dipole, deduce the Hubble parameter, and equation
(\ref{eq:errorH}) gives the error on $H(z)$ per supernova at that redshift.
We plot the relative error on $H(z)$ (which is the
same as the relative error on the dipole amplitude) in
Fig.~\ref{fig:H} for a flat universe with a cosmological constant and cold dark matter ($\Lambda$CDM)
with $\Omega_\La=0.7$. We use two values for the error on the magnitude,
$\Delta m = 0.1$ and $\Delta m = 0.15$. This is comparable to the accuracy of
recent supernova surveys like SNLS \cite{snls}.

In the future it may be possible to control
systematic errors much better -- indeed the very dipole that we want to measure is part
of the systematic error budget of current surveys \cite{hui}! Proper assessment of
the dipole contribution may therefore also help to measure the monopole
with higher precision. Our assumed value for
$\Delta m$ is probably pessimistic as most systematic uncertainties
are expected to affect the overall luminosity at a given redshift, i.e. the monopole. The
dipole which relies on the angular distribution of the luminosity should be far more
resistant to effects like for example evolution of the supernova population
with redshift.

\begin{figure}[ht]
\centerline{\epsfig{figure=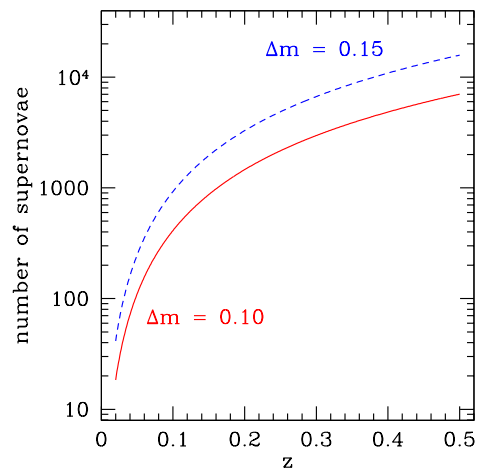,width=6.5cm}}
\caption{ \label{fig:N} We show the number of supernovae needed to measure
  $H(z)$ with an accuracy of $30$ \%, in a flat $\Lambda$CDM universe with $\Omega_\La=0.7$,
  as a function of the redshift and for two different values of the intrinsic error
  $\De m=0.1$ and $\De m=0.15$.
  }
\end{figure}

Observing $N$ independent supernovae, the mean error on the magnitude
is reduced to $\frac{\De m}{\sqrt{N}}$.  In Fig.~\ref{fig:N}
we plot the number of supernovae needed to measure
$H(z)$ with an accuracy of $30$ \%.  This number
scales quadratically with the errors; we need to measure $100$
times more supernovae to decrease the error by a factor of $10$ to
3\%. On the other hand, if we manage to decrease $\Delta m$ by a factor of $10$
through an improved understanding of supernova explosions and better
measurements, then we need $100$ times fewer supernovae.

\begin{figure}[ht]
\centerline{\epsfig{figure=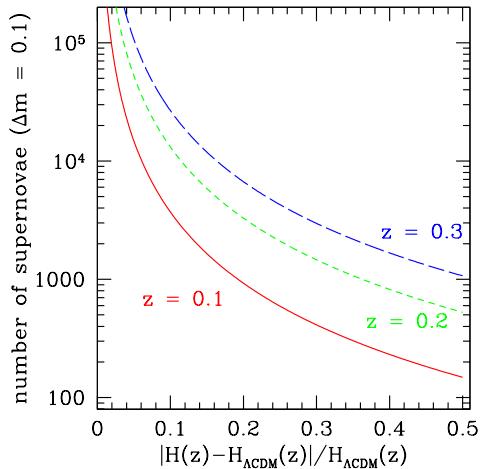,width=6.5cm}}
\caption{ \label{fig:N_H_1} We show the number of supernovae needed to
 differentiate  the measured Hubble parameter $H(z)$ from the
 theoretical one in a flat $\La$CDM universe $H_{\La CDM}(z)$, as a
 function of the relative difference
 $\frac{\left|H(z)-H_{\La\rm{CDM}}(z) \right|}{H_{\La CDM}(z)}$, for
  redshifts $z=0.1$, $z=0.2$ and $z=0.3$ and intrinsic
 error $\De m=0.1$.
  }
\end{figure}

One crucial question about dark energy is whether it does indeed behave
as a cosmological constant or not. Having measured the dipole at different
redshifts, it is possible to compare directly the measured value of $H(z)$
with the one predicted for a flat $\La$CDM universe. If the two do not
agree, dark energy must be due to a different mechanism, like
quintessence or a modification of general relativity. In Fig.~\ref{fig:N_H_1}
we plot the number of supernovae needed to distinguish the two cases,
by demanding that the difference $\left|H(z)-H_{\LCDM}(z)\right|$
be larger than the error $\De H(z)$. For comparison, the relative
difference between the Hubble parameter in a flat pure CDM universe 
and in a flat $\La$CDM universe with $\Om_\La=0.7$ is 10\% at $z=0.1$,
19\% at $z=0.2$ and 27\% at $z=0.3$. 

Our method tests directly the
expansion speed of the universe at all the redshifts where we measure
the luminosity distance dipole. Any deviation in $H(z)$ from
theoretical predictions will be immediately detected. If we measure
only the usual monopole of the luminosity distance then a well-localised 
deviation may easily be smeared out and lost by the additional integration.

Looking at the figures it is readily apparent that accuracy is best at
low redshift. This is not necessarily a drawback, as dark energy is
 expected to dominate at low redshift and so 
is best observed in this redshift range. However, below $z
\lsim 0.04$ non-linear effects probably become important
which may lead to systematic effects in the distribution of
supernovae. Future baryon oscillation
surveys will primarily target higher redshifts, $z\gsim 0.5$, where
they reach maximum sensitivity. The angular distribution of
low-redshift supernovae is therefore a complementary probe. Also the
number of supernovae needed seems quite realistic. Very large
surveys which plan to measure of the order of $10^4$ to $10^5$
supernovae are presently discussed~\cite{alpaca}.

As a final remark, even though uniform sky coverage is not essential,
a survey designed to measure the dipole should optimally cover a large
part of the sky. If we only observe supernovae in one small patch, it
may be difficult to extract the dipole without contamination from
lensing which dominates over the dipole for $\ell\gsim 100$ and $z\gsim
1$ (see~\cite{BDG}). If possible, the observed supernovae should cover
the regions of the sky aligned and 
anti-aligned with the CMB dipole where the luminosity difference is maximal.
\vspace{2mm}


In this {\em letter} we have discussed a novel method for measuring directly the
expansion history of the Universe. We have shown that the dipole of the supernova
distribution on the sky is proportional to $1/H(z)$. It is therefore possible to
extract {\em directly} $H(z)$ from the dipole. This is advantageous compared to
the monopole of the luminosity and angular diameter distance which measure only
the integral over the Hubble parameter. 
With a present data set of 44 low redshift supernovae we have measured
the dipole and it is in good agreement with the CMB.

We have also discussed the accuracy with which we can measure the
Hubble parameter, 
and found that we need a large number of supernovae. However, future planned
surveys are expected to deliver these. Given that most surveys
concentrate on high-redshift supernovae while
the dipole is most useful at moderate redshifts, $z\lsim 0.5$, it may
be preferable to propose a dedicated low-$z$ supernova survey.

Finally, the dipole is a quantity independent of the monopole. Given a survey
with a sufficient number of supernovae it is possible to measure both. This
improves the measurement of the dark energy properties and additionally serves
as a cross-check for systematic errors.

\begin{acknowledgments}

We thank Pierre Astier and Bruno Leibundgut for useful discussions. This work is supported
by the Swiss NSF.

\end{acknowledgments}


\begin{thebibliography}{99}
\bibitem{snIa}
  A. Riess et al., {\em Astronomical J.} {\bf 116}, 1009 (1998);\\
  S. Perlmutter et al. {\em Astrophys. J.} {\bf 517}, 565 (1999);
\bibitem{Blanchard} I. Maor, R. Brustein, J. McMahon and P.J. Steinhardt, Phys.
Rev. D {\bf 65}, 123003 (2002); \\
 B.A. Bassett, P.S. Corasaniti and M. Kunz, Astrophys. J. {\bf 617}, L1 (2004); \\
 M. Douspis et al.,  {\tt astro-ph/0602491}.
\bibitem{daly} R.A. Daly and S.G. Djorgovsky, {\tt astro-ph/0512576}.
\bibitem{BDG}C. Bonvin, R. Durrer and A. Gasparini, Phys Rev. {\bf
  D73}, 023523 (2006).
\bibitem{Sasaki}M. Sasaki, Mon. Not. R. Astron. Soc. {\bf 228}, 653
  (1987).
\bibitem{CMBdip} Review of part.phys., Phys. Lett. B {\bf 592}, 1 (2004).
\bibitem{riess} A.G. Riess, W.H. Press, R.P. Kirshner,
  Astrophys. J. {\bf 445}, L91 (1995).
\bibitem{snls} P. Astier et al., {\tt astro-ph/0510447}.
\bibitem{snlist}  IAU, BAT: list of supernovae,
http://cfa-www.harvard.edu/iau/lists/Supernovae.html.
\bibitem{hui} L. Hui and P.B. Greene, {\tt
  astro-ph/0512159}; \\
A. Cooray and R.R. Caldwell, {\tt
  astro-ph/0601377}.
\bibitem{alpaca} Crotts A. et al. (JEDI collaboration), {\tt astro-ph/0507043}; \\
P.S. Corasaniti et al. (ALPACA collaboration),  {\tt astro-ph/0511632}.
\end{thebibliography}
\end{document}